\documentclass[onecolumn,showpacs]{revtex4}
\usepackage{amsmath}
\usepackage{graphicx}
\usepackage{psfrag}
\usepackage[utf8]{inputenc}
\usepackage{times}

\begin{document}

\title{One-dimensional Bose-Hubbard model with pure three-body interactions}
\author{Tomasz Sowi\'nski$^{1,2}$}
\affiliation{
\mbox{$^1$Institute of Physics of the Polish Academy of Sciences, Al. Lotnik\'ow 32/46, 02-668 Warszawa, Poland}\\
\mbox{$^2$Center for Theoretical Physics of the Polish Academy of Sciences, Al. Lotnik\'ow 32/46, 02-668 Warszawa, Poland}
}
\date{\today}

\begin{abstract}
The extended Bose-Hubbard model with pure three-body local interactions is studied
using the Density Matrix Renormalization Group approach. The shapes of the first two insulating lobes are discussed, and the values of the critical tunneling for which the system undergoes the quantum phase transition from insulating to superfluid phase are predicted. It is shown that stability of insulating phases, in contrast to the standard Bose-Hubbard model, is enhanced for larger fillings. It is also shown that, on the tip of the boundary of the insulating phase, the model under consideration belongs to the Berenzinskii-Kosterlitz-Thouless universality class.
\end{abstract} 
\pacs{03.75.Lm, 67.85.Hj}
\maketitle

\section{Introduction}
Experimental progress in controlling ultra-cold atoms has opened a new chapter in our understanding of the properties of strongly-correlated many-body quantum systems \cite{Lewenstein2007,Bloch2008}. Old fashioned theoretical toy-models known from condensed matter physics are undergoing a renaissance since they provide realistic descriptions of the real quantum systems confined in optical lattices (specially arranged laser beams forming periodic potential \cite{LewensteinBook}). In the simplest case of ultra-cold bosons confined in such a potential the system is described by the Bose-Hubbard (BH) model, where single-particle tunnelings compete with local two-body interactions. The theoretical analysis of \cite{Fisher1989,Jaksch1998} shows that this competition leads directly to the phase transition from insulating phase (dominated by interactions) to superfluid phase (dominated by tunnelings). These predictions were confirmed in a spectacular experiment with ultra-cold rubidium atoms \cite{Greiner2002}. Many different extensions to the model have since been proposed and studied theoretically \cite{LewensteinBook}, and are now awaiting experimental verification.

In this article, the ground state phase diagram of a particular extension of the standard BH model is studied. Mutual interaction between particles is assumed here to be of three-body origin, i.e. these  dominate over two-body interactions. Although this assumption seems very exotic, there are some possibilities of mimicking such a model in experiments with ultra-cold atoms confined in optical lattices. In the standard description three-body terms in Hubbard-like models are introduced as an effective correction originating from interactions through higher orbitals of optical lattices \cite{Will2010}. Typically in such a scenario three-body terms are small corrections to dominant two-body terms, and they can be viewed as the first occupation-dependent correction to the on-site two-body interaction. Due to perturbative changes of single-particle wave functions, the effective three-body terms are attractive (for a repulsive gas) \cite{Will2010,Sowinski2012a}. BH models with two- and three-body interactions have been studied in many different scenarios and using different numerical techniques \cite{Sowinski2012a,Chen2008a,Zhou2010,Safavi2012,Silva2011,Al-Jib2013,Silva,Sowinski2013a,Singh,Abdullaev,Ejima2013,Sowinski2014}. Recently it was suggested that it is also possible to control three-body terms independently of the two-body ones. This can be  done by exploiting internal degrees of freedom of interacting particles \cite{Mazza2010} or via very fast dissipation processes \cite{Daley2009}. It also seems possible to control effective three-body interactions in the limit of high densities. In this limit three-body interactions can be viewed as an effective way of taking into account changes in electronic potential induced by a neighboring third particle. Typically, these changes are very small and therefore can be neglected. Nevertheless, if one tunes an external magnetic field to the value where the two-body $s$-wave scattering length vanishes, then three-body interaction induced by this mechanism dominates and in principle can be many orders of magnitude larger than two-body interaction. The consequences of a similar mechanism have been studied for the case of polar molecules interacting via long-range forces \cite{Buchler2007,Schmidt,Capogrosso2009a}. 

\section{The Model}
On this basis we now assume that two-body interactions can be neglected and the on-site energy changes only when three- or more particles are present on a given lattice site. In the one-dimensional case the Hamiltonian of the system reads:
\begin{equation}
{\cal H} = -J \sum_i \hat{a}_i^\dagger\left(\hat{a}_{i-1}+\hat{a}_{i+1}\right) + \frac{W}{6}\sum_i \hat{n}_i(\hat{n}_i-1)(\hat{n}_i-2),
\end{equation}
where $\hat{a}_i$ annihilates a boson at site $i$, and $\hat{n}_i=\hat{a}_i^\dagger\hat{a}_i$ is a local density operator. The parameter $J$ is the single-particle hopping amplitude to the neighboring site and $W$ denotes the energy cost of forming a triplet on a given lattice site. For numerical calculations, it is assumed that the lattice has $L$ sites and open boundary conditions. The properties of Hubbard-like models are strongly dependent on the average density $\rho=N/L$, where $N$ is total number of bosons confined in the lattice. For example, it is known that for models considering on-site interactions only, the insulating phase can occur only
for integer fillings \cite{Fisher1989}. Therefore, it is convenient to introduce a chemical potential $\mu$ and to rewrite the Hamiltonian in the grand canonical ensemble ${\cal K}={\cal H}-\mu\hat{N}$, where $\hat{N}=\sum_i \hat{n}_i$ is the total number of particles operator. The phase diagram of the model is described in \cite{Chen2008a,Silva}, and a similar extended BH model with non-local three-body interactions was recently studied in \cite{Capogrosso2009a}.

\section{Simple observations}
To start we investigate the properties of the system in the limit of vanishing tunneling $J\rightarrow 0$. In this limit, for any $\mu$, all correlations between neighboring sites vanish and system remains in the Mott Insulator phase (MI) with integer filling $\rho_0$. The grand canonical energy of the system is given by
\begin{equation}
{\cal E}(\rho_0,L) = L\left[\frac{W}{6}\,\rho_0(\rho_0-1)(\rho_0-2)-\mu\rho_0\right].
\end{equation}
From this relation one can easily find the boundaries of the insulating lobes (i.e. values of chemical potential for which density changes by unity). The critical values of chemical potential for which insulating phase with filling $\rho_0$ is stable are given by
\begin{equation}
\mu_\pm(\rho_0)/W = \frac{1}{2}(\rho_0 - 1)(\rho_0-1\pm 1).
\end{equation}
For any integer $\rho_0$ one finds the energy gap $\Delta(\rho_0) = \mu_+(\rho)-\mu_-(\rho_0)=W\cdot\left(\rho_0 - 1\right)$. This means that, in contrast to the standard BH model, insulating phases with larger fillings become larger. Moreover for $\rho=1$ one finds that $\mu_+(1)=\mu_-(1)=0$, i.e. the MI phase with $\rho_0=1$ does not exist at all in the system. 

\section{The phase diagram}
\begin{figure}
\includegraphics{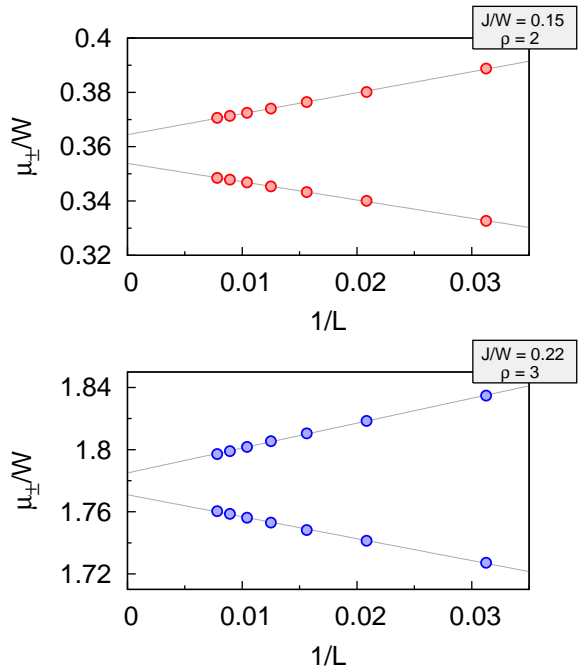}
\caption{The upper $\mu_+$ and lower $\mu_-$ boundaries of MI phase as a function of the inverse of the system size $1/L$ for two densities $\rho=2$ ($J/W=0.15$) and $\rho=3$ ($J/W=0.22$). The solid lines are linear fits to the numerical data points. Linear data extrapolation to the limit $1/L\rightarrow 0$ gives phase boundaries in the thermodynamic limit. Numerical data obtained from DMRG for $L=32,64,\dots,128$.  \label{Fig0}}
\end{figure}

To obtain the phase diagram of the studied system over the whole range of tunnelings, we follow the standard method based on energetic arguments \cite{Elesin1994}. This method is based on the observation that in the MI phase, in contrast to the SF phase, a non vanishing energy gap for adding (subtracting) a particle to (from) the system always exists. It is therefore possible to obtain the upper/lower boundary of the insulating phase with filling $\rho_0$ for given tunneling $J$, by finding numerically the ground state energy ${\cal E}_0(\rho_0,L,J)$ for $N=\rho_0\cdot L$ particles and the ground state energies ${\cal E}_{\pm}(\rho_0,L,J)$ for system with $N=\rho_0\cdot L\pm 1$ particles respectively. The upper and lower boundaries of the insulating phase are therefore given by
\begin{equation}
\mu_\pm(\rho_0,L,J) = \pm\left[{\cal E}_\pm(\rho_0,L,J)-{\cal E}_0(\rho_0,L,J)\right],
\end{equation}
as well as the energy gap within the phase
\begin{equation}
\Delta(\rho_0,L,J) = \mu_+(\rho_0,L,J)-\mu_-(\rho_0,L,J).
\end{equation}
In practice, phase boundaries obtained in this way depend strongly on the lattice size $L$. Moreover, the energy gap for the SF phase vanishes only in the thermodynamic limit $L \rightarrow \infty$, and precise localization of the phase boundaries becomes ambiguous. To overcome this problem, we perform DMRG \cite{DMRG} numerical calculations for different lattice sizes $L=32,48,\ldots,128$ and extrapolate the obtained data to the limit $L\rightarrow \infty$. This extrapolation can be done quite easily as the boundaries $\mu_\pm(\rho_0,L,J)$ treated as functions of lattice size $L$ fit almost perfectly to the linear regression with $1/L$ (for discussion see \cite{Sowinski2012a}). In Fig. \ref{Fig0} an exemplary case is presented for both $\rho_0=2$ and $\rho_0=3$. This shows the accuracy of predictions based on linear data regression to the thermodynamic limit $L\rightarrow \infty$.

\begin{figure}
\includegraphics{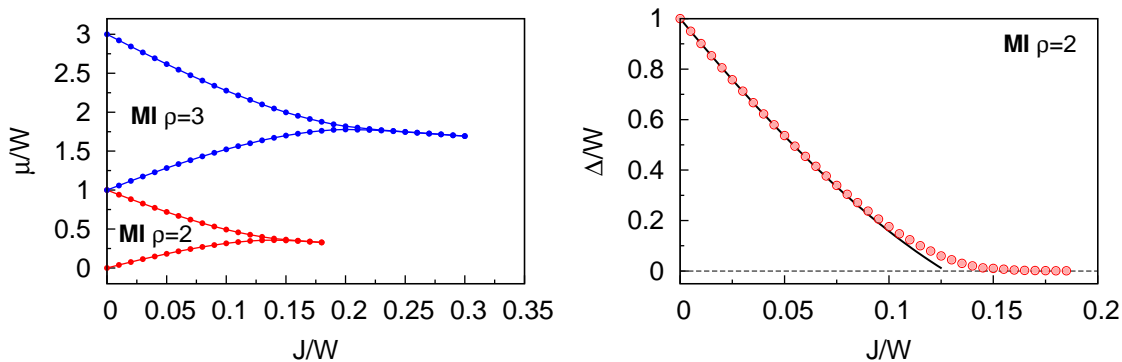}
\caption{(Left panel) The phase diagram of the Bose-Hubbard model with pure three-body interactions. In contrast to the standard Bose-Hubbard model in the first insulating lobe one finds two particles in each lattice site. Note also that the second insulating lobe for $\rho=3$ is larger than the first one for $\rho=2$. The phase diagram determined in thermodynamic limit $L\rightarrow\infty$ by extrapolating the numerical data obtained from DMRG for $L=32,64,\dots,128$. (Right panel) Rescaled single-particle gap $\Delta$ as a function of tunneling for $\rho=2$ (red circles) compared with analytical result obtained in third-order perturbation \eqref{ThirdOrderGap} (solid black line). \label{Fig1}}
\end{figure}

Finally, the phase diagram of the system is obtained by plotting extrapolated values of $\mu_\pm(\rho_0,L\rightarrow\infty,J)$ as functions of tunneling (Fig. \ref{Fig1}). The result is consistent with previous analytical predictions in the limit of vanishing tunneling. The second insulating lobe ($\rho=3$) is broader than the first one ($\rho=2$) in the direction of chemical potential as well as in the direction of tunneling. This means that, in contrast to the standard Bose-Hubbard model, the critical tunneling $J_c$ for which system undergoes the phase transition from MI to SF phase is shifted to larger values for higher fillings.

From the numerical point of view the most problematic part of these calculations lies in determining the critical tunneling $J_c$ for which the system undergoes phase transition from MI to SF phase. Theoretically, this point is defined as a tunneling for which the energy gap $\Delta(J)= \mu_+(J)-\mu_-(J)$, calculated in the thermodynamic limit $L\rightarrow\infty$, vanishes. Unfortunately, due to the numerical errors, this definition can not be adopted directly. The phase diagram obtained above allows us to estimate the critical tunneling $J_c/W\sim 0.19$ for $\rho=2$ and $J_c/W\sim 0.28$ for $\rho=3$. 

At this point it is worth comparing the energy gap $\Delta(J)$ obtained numerically to the analytical results obtained recently in \cite{Ejima2013}. In that paper the authors perform perturbative calculations for a general BH model with two- and three-body local interactions (for $\rho=2$). In the third-order of perturbation with respect to the tunneling, in the particular case of vanishing two-body interactions, the result reduces to the form

\begin{equation} \label{ThirdOrderGap}
\frac{\Delta^{(3)}(J)}{W} = 1 - 10\frac{J}{W}+\frac{38}{3}\left(\frac{J}{W}\right)^2+\frac{116}{3}\left(\frac{J}{W}\right)^3+\ldots
\end{equation}

As it is seen in the right panel of Fig. \ref{Fig1}, the energy gap obtained numerically fits almost perfectly to the predictions of \eqref{ThirdOrderGap}. The deviations are clearly visible for larger tunnelings where the third-order approximation breaks down.

\section{Berenzinskii-Kosterlitz-Thouless transition}
\begin{figure}
\includegraphics{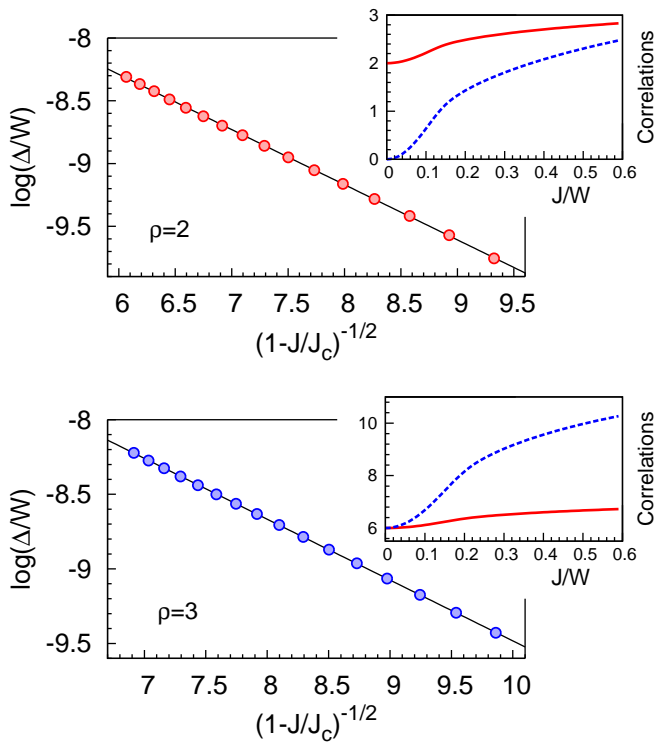}
\caption{Energy gap of the insulating lobe $\Delta$ as a function of tunneling $J$ for two integer fillings $\rho_0=2$ (upper panel) and $\rho_0=3$ (bottom panel). With this scaling the numerical points fit to the linear behavior predicted by the Kosterlitz-Thouless universality class \eqref{GapRel}. This suggests that studied model belongs to the same universality class as standard Bose-Hubbard model. Numerical data obtained from DMRG for $L=128$. In the insets the correlation functions ${\cal C}_2$ (red solid line) and ${\cal C}_3$ (blue dashed line) as functions of tunneling $J/W$ are presented. \label{Fig2}}
\end{figure}

In order to determine the critical tunneling more precisely two independent but complementary methods may be used. The first is based on the assumption that near the critical point the studied system belongs to the same universality class as the standard Bose-Hubbard model. At the phase transition the standard BH model in $d$ dimensions can be mapped to the $d+1$-dimensional XY model. Therefore, in the one-dimensional case the phase transition belongs to the Berenzinskii-Kosterlitz-Thouless class (BKT)\cite{Berenzinskii1972,Kosterlitz1973}. As was shown recently, the universality class does not change when one extends the standard BH model with local three-body terms \cite{Sowinski2012a}. This suggests, that even in the limit of vanishing two-body terms (as studied here) the universality class can remain unchanged. If true, this is a way to obtain the critical tunneling $J_c$. Indeed, for the BKT transition the energy gap $\Delta(J)$ in the vicinity of the critical tunneling $J_c$, vanishes as
\begin{equation} \label{GapRel}
\Delta(J) \sim \exp\left[-\frac{\alpha}{\sqrt{1-J/J_c}}\right].
\end{equation}
Therefore, if the critical tunneling $J_c$ was known and indeed the relation \eqref{GapRel} would hold, then by plotting $\log\,\Delta(J)$ against $\sqrt{1-J/J_c}$ the data points should follow a linear regression. 
Moreover, this can happen only for a unique value of $J_c$ and, due to uniques of the relation \eqref{GapRel}, only if the transition is of BKT type. Plots in Fig. \ref{Fig2}  show that the BKT scaling is satisfied with an appropriately chosen critical value of the tunneling $J_c$. In this way we confirm that the phase transition is indeed of BKT type. Values of critical tunneling obtained in this way are $J_c/W=0.191 (\pm 0.005)$ and $J_c/W=0.282 (\pm 0.005)$ for $\rho_0=2$ and $\rho_0=3$ respectively. Uncertainties in the critical tunnelings may be estimated from comparison of the results obtained for different system sizes $L=118,\ldots,128$. In all these cases the critical tunneling differs from estimated values by no more than estimated uncertainties.  

For completeness  local two-body ${\cal C}_2 = \langle \hat{a}_m^{\dagger 2}\hat{a}_m^2\rangle$ and local three-body ${\cal C}_3 = \langle \hat{a}_m^{\dagger 3}\hat{a}_m^3\rangle$ correlation functions (for the middle lattice site $m=L/2$) are plotted in the insets of Fig. \ref{Fig2}. For both fillings studied ($\rho_0=2$ and $\rho_0=3$), in the vicinity of the phase transition the three-body correlation ${\cal C}_3$ changes its behavior, which can be viewed as a changing of ground-state properties. Note however that in the limit of large tunneling, both correlation functions necessarily approach the values of the standard BH model. 

\section{Entanglement entropy approach}
\begin{figure}
\includegraphics{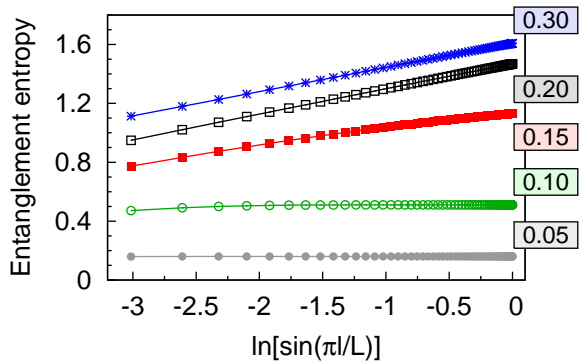}
\caption{Entanglement entropy of the subchain of length $l$ for a number of example tunnelings $J/W$ ($\rho=2$). With chosen scaling the numerical points fit to the linear predictions of CFT. In the MI phase (low tunnelings) the slope of the corresponding lines (proportional to the central charge $\mathtt c$) is equal to 0. In the SF phase (large tunnelings) the line gradients saturate on the value $\sim 1/6$. This corresponds to the central charge value ${\mathtt c}=1$ predicted by the Tomonaga-Luttinger liquid theory. Numerical data obtained from DMRG method for $L=128$. \label{Fig3}}
\end{figure}

The phase transition from the MI to SF phase can be also identified using a complementary method, by looking for changes in the behavior of the entanglement entropy (EE) of the subsystem ${\cal S}(l,L) = -\mathrm{Tr} \left[\hat\rho_l\,\mathrm{ln}\hat\rho_l\right]$. Here, $\hat\rho_l =\mathrm{Tr}_{L-l} |\mathtt{G}\rangle\langle\mathtt{G}|$ is the reduced density matrix of the subchain of length $l$ obtained by tracing-out remaining degrees of freedom from the ground state of the system $|\mathtt{G}\rangle$. The scaling behavior of the EE is well known in the thermodynamic limit, i.e. when $L\rightarrow \infty$. In the SF phase, due to the nonlocal correlations in the system, EE treated as a function of size of the subsystem is logarithmically divergent with $l$. In contrast, in the MI phase, long-range correlations vanish and therefore entanglement entropy saturates for large enough subsystem sizes $l$. These facts have some consequences also for finite size $L$ of the full system. As predicted by conformal field theory, depending on the boundary conditions, in the SF phase entanglement entropy is the following function function of $l$ \cite{Calabrese2004,Laflorencie2006}
\begin{equation} \label{EntropyEq}
{\cal S}(l,L) = \frac{\mathtt{c}}{3\kappa}\,\mathrm{ln}\left[\frac{\kappa L}{\pi}\sin\left(\frac{\pi l}{L}\right)\right] + s(L) + {\cal O}\left(\frac{l}{L}\right).
\end{equation}
The parameter $\kappa$ depends on the boundary conditions and  is equal to $1$ or $2$ for periodic or open boundary conditions respectively. The pre-factor ${\mathtt c}$ is related to the central charge of corresponding conformal field theory. For non-critical phases (like MI phase) it is zero, whereas it is non zero whenever the system manifests some non-local correlations. It is known that deep in the SF phase, due to the equivalence with Tomonaga-Luttinger liquid \cite{Cazalilla2011}, central charge ${\mathtt c}=1$. To show that EE in the studied system can be well understood with this description, Fig. \ref{Fig3} plots entanglement entropy ${\cal S}(l,L)$ as a function of scaled subsystem size $\mathrm{ln}\left[\sin\left(\frac{\pi l}{L}\right)\right]$ obtained from DMRG calculations with $L=128$ for different tunnelings $J/W$ and $\rho_0=2$. With appropriate scaling the numerical points fit almost perfectly to lines which is in agreement with the predictions of \eqref{EntropyEq}. The gradients of these lines are directly related to the central charge of the many-body quantum state. 

\begin{figure}
\includegraphics{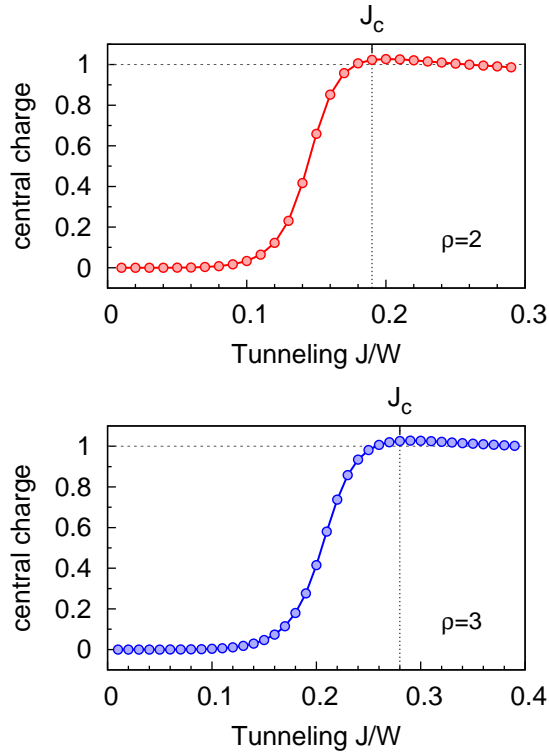}
\caption{Central charge $\mathtt c$ as a function of the tunneling rate $J$ for two integer fillings $\rho=2$ (upper panel) and $\rho=3$ (bottom panel) determined from the behavior of the EE \eqref{EntropyEq}. For small tunnelings in MI phase the central charge is equal to $0$ and in deep SF phase it is equal to $1$ in accordance with Luttinger Liquid theory. Near the quantum phase transition we observe a rapid change of central charge, and for critical tunneling $J_c$ the central charge $\mathtt c$ achieves the maximal value. The value of the critical tunneling $J_c$ agrees with the value determined from decaying of the energy gap of insulating lobe \eqref{GapRel}. Numerical data obtained from DMRG method for $L=128$.   \label{Fig4}}
\end{figure}

The method described above enables one to plot the central charge $\mathtt c$ as a function of tunneling $J$. The results for two integer fillings $\rho_0=2$ and $\rho_0=3$ are presented in Fig. \ref{Fig4}. In both cases, in the MI phase the central charge vanishes and deep in the SF phase it saturates at the expected value ${\mathtt c}=1$. For moderate values of tunneling rapid change in the behavior of entanglement entropy is observed. The central charge achieves its maximal value at the critical point predicted with the previous method. Such behavior of the central charge is very similar to the situation observed in the standard BH model \cite{Ejima2012}. It is believed that non monotonicity in the central charge behavior is a direct consequence of the finite size of the system, and in the thermodynamic limit it smoothly flows to  ''step-like'' behavior. The maximal value of the central charge $\mathtt c$ obtained from finite size calculations is reached in the neighborhood of the critical tunneling $J_c$. All numerical results obtained here fully agree with all these properties.

\section{Conclusions}
The phase diagram for the one-dimensional extended Bose-Hubbard model with pure three-body interactions was studied. It was shown that insulating lobes are present for integer fillings $\rho_0\geq 2$ and that their shapes, in contrast to the standard BH model, become larger for larger $\rho_0$. Three-body interactions lead to enhanced stability of the MI phase in the $\mu-J$ phase diagram. The first two MI lobes were discussed in details with DMRG calculations for different system sizes. Values of critical tunnelings $J_c$ for which the system undergoes phase transition from MI to SF were determined. It was also shown that the studied model belongs to the BKT universality class in analogy to the standard BH model. 

\section{Acknowledgements}
The author thanks R. W. Chhajlany, P. Deuar, M. Gajda, and M. Lewenstein for their fruitful comments and suggestions. This research was supported by the (Polish) National Science Center Grant No. DEC-
2011/01/D/ST2/02019. The author acknowledges support from the Foundation for Polish Science (KOLUMB
Programme; KOL/7/2012) and hospitality from ICFO.


\begin{thebibliography}{99}
\bibitem[{DMR()}]{DMRG}
\bibinfo{note}{Here I use the finite-size DMRG algorithm. The number of bosons
  occupying a given lattice site is limited to 5. In each step of the DMRG
  calculations no more than 600 states in the Schmidt-decomposition are stored.
  Convergence of the method is checked for each numerical point, and the
  relative error for the energy of the final state is smaller than $10^{-8}$.}
\end{thebibliography}
\end{document}